\documentclass[letterpaper,journal]{IEEEtran}
\renewcommand\IEEEkeywordsname{Keywords}
\usepackage{cite}
\usepackage[cmex10]{amsmath}
\usepackage{algorithm}
\usepackage{algorithmic}
\usepackage[normalem]{ulem}

\usepackage{graphicx}
\usepackage[T1]{fontenc}
\usepackage[tight]{subfigure}
\usepackage{geometry}

\newcommand{\searsa}{SEARS\ }

\newcommand{\sears}{SEARS}

\newcommand{\kguodone}[1] { }

\begin{document}

%\title{Space Efficient and Fault Tolerant Cloud Storage System with Fast File Retrieval: Deduplication with Erasure Coding}
\title{SEARS: Space Efficient And Reliable Storage System in the Cloud}

\author{ 
$^\dagger$Ying Li, $^\ddagger$Katherine Guo, $^\dagger$Xin Wang, $^\ddagger$Emina Soljanin, $^\ddagger$Thomas Woo
\\ $^\dagger$Dept of Electrical and Computer Engineering, Stony Brook University
$^\ddagger$Bell Labs, Alcatel-Lucent
\thanks{\copyright 2015 IEEE. This paper has been accepted to LCN 2015. Personal use of this material is permitted. Permission from IEEE must be obtained for all other uses, including reprinting/republishing this material for
advertising or promotional purposes, collecting new collected works for resale or
redistribution to servers or lists, or reuse of any copyrighted component of this work in other
works.}}

\maketitle

%we can make this paper suitable for both backup batch processing and real time online system. chunk-level binding maybe suitable for batch processing and user-level binding suits real time case. in the simulation, we can compare space efficiency between chunk level binding with other batch processing schemes, and compare the speed between user level with other online systems.

\begin{abstract}

Today's cloud storage services
must
%are required to
offer storage
reliability and fast data retrieval for large amount of data without
sacrificing storage cost.  We present SEARS, a cloud-based storage
system which integrates erasure coding and data deduplication to
support efficient and reliable data storage with fast user response
time.  With proper association of data to storage server clusters,
SEARS provides
%the flexibility to mix
flexible mixing of different configurations,
%making it
suitable for
%both
real-time and archival applications.

Our prototype implementation of
SEARS over Amazon EC2 shows that it outperforms existing storage
systems in storage efficiency and file retrieval time.
For 3 MB files, \searsa delivers retrieval time of $2.5$ s
compared to $7$ s with existing systems.

\end{abstract}

\vspace{\baselineskip}

\section{Introduction}
\label{sec:intro}

Data from connected devices today are flowing into data centers with an
unprecedented rate.
%Survey results indicate that
More than half of
the companies in the survey of global enterprise market currently
store at least 100 TB of data and one-third expect their data to
double in the next two to three years~\cite{ms-survey}.

The cloud infrastructure enables low-cost and scalable file storage
that provides global file access. Any file system must offer reliable
storage whether through file duplication that requires more space but
less computation complexity such as GFS~\cite{GFS} or through erasure
coding that requires less space but more computation complexity such
as RAID systems~\cite{RAID}.
At the same time, raw data exhibit redundancy across files. This redundancy can
be explored to reduce storage cost mainly for backup systems
~\cite{LBFS,ADB,SIL}. Using these techniques, data are
divided into chunks and unique data chunks are stored once and
referenced multiple times.
Different from archival systems, cloud-based storage systems are required
to support interactive user access with reasonable response time.

%A user data retrieval process generally consists of two parts: the
%download or transfer of data over the Internet and the
%ecode or data reconstruction operation.

We propose a cloud-based file system named \textbf{SEARS}-Space Efficient
And Reliable Storage system that exploits the deduplication technique to reduce storage and traffic
cost as well as the erasure coding technique to increase both the data reliability and the file retrieval speed.
Given a file, there are different ways to associate data chunks with
available storage servers and retrieve data.  Archive-based backup
systems mainly care about storage efficiency and
reliability. However, interactive cloud storage systems also care about
file retrieval speed.  To meet different application needs, we propose
two data-server binding schemes with different performance goals: (1)
faster file access speed or (2) higher storage efficiency.

We aim for SEARS to serve as a reference
%for the
design
%of
for a flexible cloud storage framework that can support
customized level of deduplication,
%allow the selection of modes of
modes of coding and server binding,
%and enable the mix of
and the mix of different modes.
%The ultimate flexibility of the platform would help to gracefully handle
Its flexibility handles
different application scenarios, from batch-centric archival to real-time.

\textbf{ Related Work:}
Recent studies have reported that erasure coding can guarantee the
same level of content accessibility with lower storage than
replication~\cite{berk_isit, ulric_muriel_emina,GYE_JSAC}.

File deduplication relies 
%on the central idea of 
dividing files into chunks and eliminating the need to store
or transfer identical chunks multiple times.  
%Various techniques have been studied to reduce the storage and network bandwidth requirement for client server synchronization in a storage system. 
LBFS~\cite{LBFS} introduced content based chunking with Rabin
fingerprints~\cite{rabin}.  Various work improves on the idea by
compressing data chunks~\cite{ADB}, comparing chunks belonging to
highly related files~\cite{SIL}, switching between large and small
chunk sizes to discover more overlapping regions~\cite{FBC}.

The tradeoff between increasing data reliability and deduplicating data
poses great challenge to the design of a reliable and space efficient storage system.
%
%For systems without using coding~\cite{Bha-Pro-2006, Dub-Hyd-2009}, each chunk %is assigned a different redundancy
%level based on the chunk popularity. It takes extra time to record and check ch%unks to classify their redundancy levels, with chunks still being duplicated al%beit at different levels.
%
The archival storage system R-ADMAD~\cite{Liu-Rad-2009} combines the
content-based chunk deduplication with erasure coding. 
%to achieve reliability and space efficiency.
%To compensate for the inefficiency in the processes of coding and chunk-server association, 
However, it uses large and fixed-size data chunk containers,
which 
%compromises
reduces deduplication efficiency and 
%also 
makes it 
%not suitable 
ill-suited for delay-sensitive 
%cloud 
applications.

%variable sized chunks, followed by packing variable-length chunks into
%fixed-sized objects, followed by applying erasure code to bigger
%fixed-sized objects.
%called ``chunk containers''. However,
%All chunks in the system are considered equally important, and further coded with
%the same erasure code with the same redundancy level.
%This is unavoidable as erasure coding is applied at the chunk container level,
%therefore all chunks within the same chunk container can only be
%offered the same reliability level.
%

%For erasure coded storage systems, a systematic analysis of chunk
%reliability is presented in~\cite{Li-Rel-2010} and~\cite{Nam-Rel-2011}, where
%the concepts of ``redundancy specification'' and ``chunk loss severity''
%are studied in~\cite{Li-Rel-2010} and~\cite{Nam-Rel-2011} respectively.

%\input Sec2_Related
\section{SEARS Architecture}
\label{sec:system}

%\begin{figure}[t]
\begin{figure}[htbp]
\centering
\includegraphics[width=3.0in]{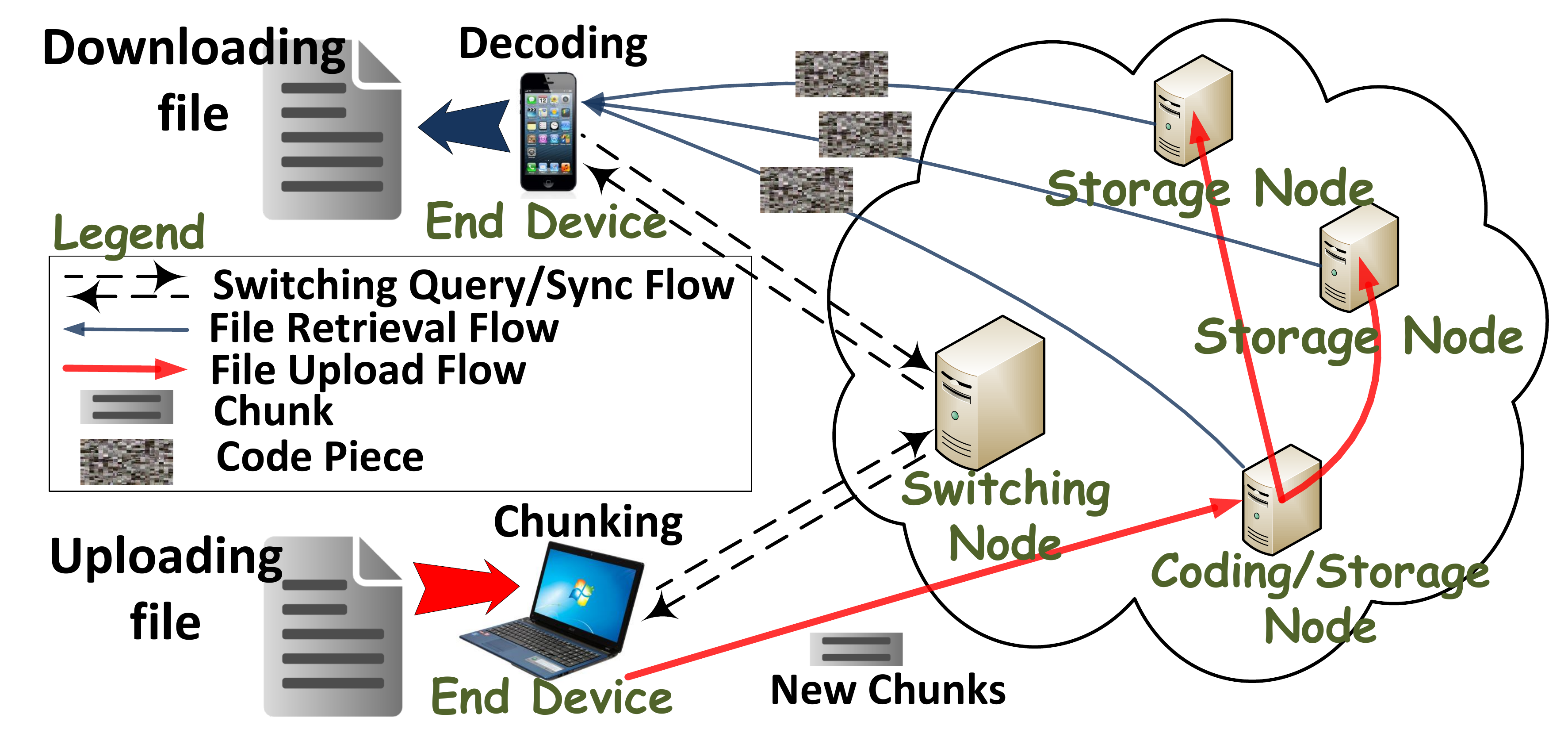}
% where an .eps filename suffix will be assumed under latex,
% and a .pdf suffix will be assumed for pdflatex; or what has been declared
% via \DeclareGraphicsExtensions.
\caption{\searsa system overview. One end device (laptop) uploads a
  file where the file is chunked at end device and the meta-data for
  the file is uploaded to the \textbf{switching node} for the
  user. Unique chunks for the file missing from \searsa are sent to and coded at
  the \textbf{coding node} which is one of the server nodes in the
  cluster storing code pieces of the chunks for the file.  Another end
  device (smart phone) downloads a file where code pieces of each
  unique chunk are retrieved from multiple storage nodes in \searsa
  concurrently.}
\label{fig:system}
%\vspace{-0.12in}
\end{figure}

Figure~\ref{fig:system}
%depicts
shows
the \searsa system architecture
consisting of storage server nodes operating in a data center.
%
%Users access \searsa through various access channels (e.g., Wired LAN,
%Wireless LAN, or Cellular network) and different end devices (e.g.,
%PC, tablet, laptop or smart phone).
%
Users use \searsa as any file system by storing (or uploading) files
to server nodes; and retrieving (or downloading) files from server
nodes.  Each user accesses \searsa through a designated storage server
node we call \textbf{switching node} for the user and all the user's
files.
%
%\footnote{Note that to better support user mobility, multiple
%  switching nodes can be used for a user, with each node closest to
%  the access network the user is using to access
%  \sears. Synchronization between these switching nodes are required
%  in the background to tradeoff extra space required for user file
%  meta-data and bandwidth used for synchronization to reduce file
%  access bandwidth usage in the network and improves file access
%  response time.}.
%
Each user end device is configured with host name or the IP address of
the the user's switching node since it is the first node to reach
\sears.
We consider a total of $N$ nodes in \searsa
divided into
%, and divide them into
non-overlapping \textbf{clusters} of size $n$.
The reason of forming cluster of nodes
is due to the need of storing coded chunks at multiple nodes for reliability. We assign each
cluster with a unique
%id called
\textbf{cluster id}.
We focus on the single data center configuration in this
work. However, the concept of \searsa can be naturally extended to
multiple data centers.

%\subsection{Content-based Chunking Operation}
%\label{subsec:chunking}
\textbf{Content-based Chunking Operation:}
Before storing data,
\searsa first removes redundant content.
%ahead of data storage.
Files are divided into chunks and unique chunks are stored only once.
We use content-based chunking
to better capture redundancy~\cite{ProbabilisticDedup}.
Using smaller chunk size can
result in more duplicate chunks thus achieving higher levels of
deduplication. However, it
also results in larger number of chunks and therefore larger overhead
in meta-data management and reduced system performance.
Furthermore, disk operations benefit from continuous data access,
while smaller chunks lead to less efficient random access pattern.
To balance the tradeoff, we choose average chunk size of $4$
KB~\cite{DtaRoutingDedupCluster}\cite{LBFS} and enforce the minimum
and maximum chunk sizes to be $1$ KB and $8$ KB respectively.
For each chunk, we apply the 160-bit SHA-1 hash function~\cite{DtaRoutingDedupCluster}
to generate a fixed-size hash value to serve as the
\textbf{chunk id}.

%\subsection{File Storage Operation}
%\label{subsec:upload}
\textbf{File Storage Operation:}
Ahead of data storage, \searsa explores both intra-file and inter-file
content redundancy and eliminates all redundant content.
In the first step, \searsa eliminates intra-file redundancy as follows.  Before a
user file is uploaded into \sears, the end device applies
content-based chunking to the file, and generates
%a unique identifier
%we call \textbf{chunk id}
chunk id for each chunk,
%\footnote{We design the file
%  chunking process on the user's end device. However, for a mobile
%  device, this process can be offloaded into a device separate from
%  \searsa with more processing capability than the mobile device. An
%  example of such a device can be a Cloudlet proxy
%  server~\cite{cloudlet09}.},
and produces \textbf{file chunk-meta-data} for the file, which is composed of
a sequence of
entries for all chunks in the file and each entry consists of a chunk
id and a cluster id specifying the cluster that stores the chunk.
The file
chunk-meta-data is stored at (1) the user's end device and uploaded to
(2) the \searsa switching node serving the user.  After this process, 
only non-repeating chunks will be kept so that
intra-file redundancy can be eliminated.

% Data Organization section is here.
A file in \searsa is represented by its file chunk-meta-data.
Each unique chunk is stored as $n$ code pieces in an $n$-node cluster.
The user's switching node keeps a \textbf{chunk-meta-data-table} that
stores one file chunk-meta-data for each file belonging to the user.
As a chunk can appear in multiple files, we define the
\textbf{reference count} for a chunk as the number of files in \searsa
that the chunk appears in.
The chunk reference count is updated as \searsa evolves with file
addition, removal and update.

%Next,
In the second step, \searsa eliminates inter-file redundancy across the set of nodes
responsible for storing the file as follows.
%\footnote{The set of \searsa
%  nodes storing chunks of a file are determined by the binding schemes
%  in Section~\ref{sec:binding}.}.
The user's switching node in \searsa
%then
removes
chunk ids already in the set of nodes and forms a list of ids of
missing chunks for the end device to upload directly to the set of storage nodes.
This means only unique
chunks that are not present in the set of \searsa nodes are uploaded
from the user's end device. As a result, bandwidth between the user's end device
and \searsa is only required to transfer
%necessary and
non-redundant
data.

%\subsection{File Retrieval Operation}
%\label{subsec:download}
\textbf{File Retrieval Operation:}
Whenever an end device retrieves a file from \searsa for the first time,
%for instance, the file was originally uploaded to \searsa from another
%end device or another storage system,
the requesting end device does
not have the file chunk-meta-data and the retrieval request is sent to
the user's switching node. The switching node first sends back the
file chunk-meta-data. The end device then checks the list of chunk ids
in the file chunk-meta-data against the list of chunk ids already in
its local storage, and determines the missing chunks
%required to reconstitute
needed to construct the file. The end device then only requests the missing
chunks from \sears.

%\subsection{File Chunk-Meta-Data Synchronization Operation}
%\label{subsec:synch}
\textbf{File Chunk-Meta-Data Synchronization Operation:}
In the case when the end device and its responsible switching node in \sears cloud
each has a version of the file chunk-meta-data,
synchronization is required to resolve any conflicts. We
follow the policy
%that when the file's chunk-meta-data exist at both locations,
for the copy with the latest time-stamp
%overwrites
to overwrite the one
with an earlier time-stamp.  We assume clock synchronization between
the user's end device and \searsa is provided with mechanisms such as
NTP~\cite{NTP}.

%\subsection{Erasure Coding and Decoding Process}
%\label{subsec:coding}
\textbf{Erasure Coding and Decoding Process:}
In \sears, each unique chunk first reaches a node in the cluster that
stores the code pieces of the chunk, we call \textbf{coding node}.
The coding node then divides the chunk into $k$ equal-sized pieces and
codes it into $n$ code pieces through $(n,k)$ erasure coding with $n
\ge k$. These $n$ code pieces are associated with a cluster of $n$
storage nodes and exactly one piece is stored in one node in the
cluster. Note that any node in the cluster can serve as the coding
node for a chunk to be stored at the cluster.
%
%
%We present the suite of algorithms to associate
%a chunk to a cluster in Section~\ref{sec:binding}.
%

Whenever the user's end device requests a missing chunk in a file based on
the file chunk-meta-data, it issues $n$ concurrent requests to the $n$
nodes in the cluster identified by the cluster id and as soon as $k$
code pieces are received, it reconstructs the chunk and terminates any
ongoing connection to the remaining $n-k$ nodes.
This design benefits from parallel download of data to reduce \searsa
response time as we show in Section~\ref{sec:perf}.

%\subsection{Data Organization}
%\label{subsec:data}
%A file in \searsa is represented by its file chunk-meta-data.
%Each unique chunk is stored as $n$ code pieces in an $n$-node cluster.
%The user's switching node keeps a \textbf{chunk-meta-data-table} that
%stores one file chunk-meta-data for each file belonging to the user.
%As a chunk can appear in multiple files, we define the
%\textbf{reference count} for a chunk as the number of files in \searsa
%that the chunk appears in.
%The chunk reference count is updated as \searsa evolves with file
%addition, removal and update.

\section{Server Binding Schemes}
\label{sec:binding}
%\begin{figure}[htbp]
%\centering
%\includegraphics[width=2.3in]{UserLevelBinding.pdf}
%% where an .eps filename suffix will be assumed under latex,
%% and a .pdf suffix will be assumed for pdflatex; or what has been declared
%% via \DeclareGraphicsExtensions.
%\caption{Illustration of User Level Binding rule.}
%\label{fig:UserLevelBinding}
%%\vspace{-0.12in}
%\end{figure}
%
%
%
%\begin{figure}[h]
%\centering
%\includegraphics[width=2.3in]{ChunkLevelBinding.pdf}
%% where an .eps filename suffix will be assumed under latex,
%% and a .pdf suffix will be assumed for pdflatex; or what has been declared
%% via \DeclareGraphicsExtensions.
%\caption{Illustration of Chunk Level Binding rule.}
%\label{fig:ChunkLevelBinding}
%%\vspace{-0.12in}
%\end{figure}
%
%
%
%
%
%\begin{figure}[h]
%\centering \includegraphics[width=2.6in]{FileLevelBinding.pdf}
%% where an .eps filename suffix will be assumed under latex,
%% and a .pdf suffix will be assumed for pdflatex; or what has been declared
%% via \DeclareGraphicsExtensions.
%\caption{Illustration of File Level Binding rule.}
%\label{fig:FileLevelBinding}
%%\vspace{-0.12in}
%\end{figure}

%\begin{figure*}[ht]%
\begin{figure}[htbp]%
\subfigure[Chunk Level Binding (CLB)]{\includegraphics[width=0.5\linewidth]{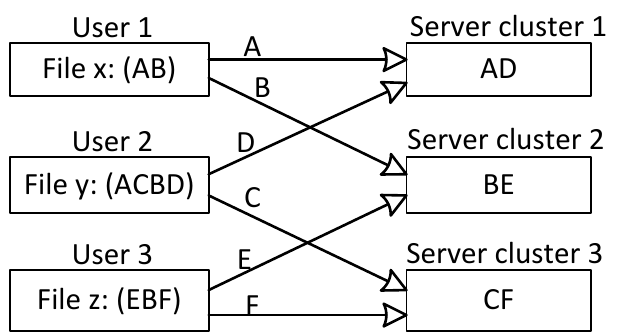}\label{fig:CLB}}%
\subfigure[User Level Binding (ULB)]{\includegraphics[width=0.5\linewidth]{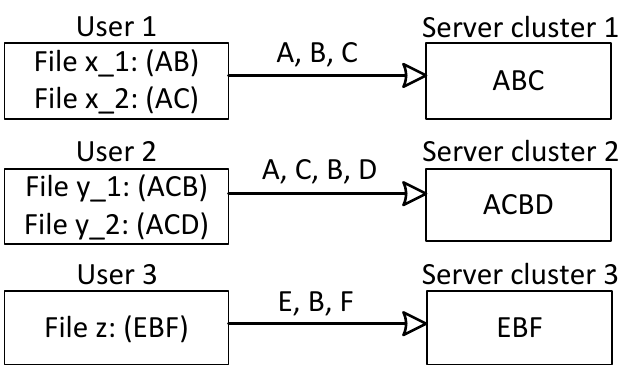}\label{fig:ULB}}%
\caption{Illustration of the binding schemes.}%
\label{fig:binding}
\end{figure}

%\subfigure[File Level Binding (FLB)]{\includegraphics[width=0.33\linewidth]{Fil%eLevelBinding.pdf}\label{fig:FLB}}%

%We consider \searsa with 
Consider \searsa nodes grouped into $M$ clusters of size
$n$.  A file to be stored in \searsa is divided into chunks and each
chunk is coded into $n$ code pieces to be stored in a cluster. A key
design question for \searsa is to determine how to associate data to
clusters. We call this the \textbf{binding} process.
Different applications have different requirements for cloud-based
storage services, including fast file retrieval, small space usage in
order to reduce storage cost.
%and balanced performance between the
%two. Accordingly, we design three different 
%
We design binding schemes across the
spectrum of application requirements 
namely Chunk Level Binding and User Level 
Binding with examples in Figure~\ref{fig:CLB} and~\ref{fig:ULB}
respectively.
%and File Level Binding with examples in Figure~\ref{fig:ULB},~\ref{fig:CLB}
%and~\ref{fig:FLB} respectively.

%\subsection{Chunk Level Binding (CLB)}
%\label{subsec:ULB}

\textbf{Chunk Level Binding (CLB):}
%At the other end of the spectrum are 
For archival applications that runs in the background and demands
storage efficiency, the binding process must offer system wide data
deduplication.  The \textbf{Chunk Level Binding (CLB)} scheme selects
the best cluster to store each chunk. CLB is ideal for large media
content repository like YouTube and NetFlix where users share the same
or similar content.
Each unique chunk entering \searsa is assigned to
a cluster
such that storage space of all clusters are evenly consumed as time passes.
Note that all storage and retrieval requests must pass through the
user's switching node.  
%
%Further more, each cluster keeps Chunk-BF in
%its nodes. With CLB, each cluster periodically sends its Chunk-BF to
%the user's switching node so that the switching node can determine
%with certainty if a chunk is not stored in a cluster, and if it is
%possible that a chunk is stored in a cluster, the chunk id must be
%sent to the cluster to check against its Chunk-Addr-HT. After these checks,
%the switching node can instruct the end device to store a chunk to a cluster or
%retrieve a chunk from a cluster.  
%
To distribute load evenly to
clusters, we use a greedy algorithm to assign a chunk to the cluster
with the largest amount of free storage space.

%In summary, CLB ensures that each chunk is stored only once
%in \sears, and chunks are evenly distributed among all clusters, thus
%all inter-user and intra-user redundancies at the chunk level are
%eliminated, and maximum storage saving is achieved.  However, due to
%its complex chunk location recording and inquiring process, more space
%is required to store file chunk-meta-data and user response time is higher
%than ULB as we show in Section~\ref{sec:perf}.

%\subsection{User Level Binding (ULB)}
%\label{subsec:ULB}
\textbf{User Level Binding (ULB):}
For interactive applications with emphasis
on  promptness of file retrieval, the binding scheme must offer
simplicity in chunk retrieval.
%location and retrieval.
%should simplify the complexity in locating
%necessary chunks, rather than going through complicated lookup
%process for each chunk.
%Intuitively,
The \textbf{User Level Binding (ULB)} scheme
binds each user with a fixed cluster
and simplifies file retrieval process as all chunks of this user are stored
in the same cluster.
%
%A straightforward design is a static binding where once a user is
%bound with a cluster, the association does not change over time
%until the cluster runs out of space. The binding happens when a user
%first enters the system, and is determined based on the current
%storage space utilization of each cluster. ULB balances load across
%all \searsa nodes, and assigns users evenly across all clusters. 
Initially each user is assigned a fixed cluster. When
storage capacity is exhausted at the cluster assigned for the user, a
new cluster is assigned to future files from the user. This is
equivalent to assigning a subset of user files to a separate user and
only intra-set redundancy within the subset of files can be captured.
%
%In summary,
ULB incurs at most one extra cluster id for a subset of user files,
offers simple retrieval process but
sacrifices space efficiency, as the
chunks stored in different clusters belonging to different users (or
even the same user) can not be exploited globally during the
deduplication process.

%\subsection{A Flexible Platform}

The two binding schemes described so far offer different tradeoffs
in space saving and file retrieval response time.
However, they are just examples to showcase the
flexibility in the design of \searsa. We design \searsa to be a powerful platform that use both
deduplication and erasure coding in the best combination to fit various
application needs.

%\input Sec3_SystemOverview
%\input Sec4_MainScheme
%\input Sec5_Simulation
%For the VM-disk-image restore evaluations, we downloaded 62 pre-made VM disk
%images from VMware's virtual appliance marketplace8 , Bagvapp's Virtual Appliances 9, a nd Thoughtpolice 10 which different guest OSs, and versions, such as CentOS, Debian, Fedora, FreeBSD, OpenSUSE, and Ubuntu. The total size is about
%178 GB, and the deduplication ratio is 48.9% with 4KB chunk size. In the VM-diskimage
%restore evaluations, we also use the iodump utility to collect disk I/O operations

\section{Performance Evaluation}
\label{sec:perf}
We evaluate the performance of
our prototype implementation of \searsa over Amazon EC2~\cite{EC2}.
%
%\subsection{\textbf{Data Traces}}
%\label{subsec:trace}
%
%To represent general archival applications, we use data
%downloaded from The Internet Archive
%(http://www.archive.org), of which three type of data traces (text, video and s%oftware) has been tested as in Table~\ref{tab:dataset}.
%
%
%\begin{table}[htb] %seems putting it at bottom saves more space
%$small
%\caption{}
%\subtable[Archival Data Set]{
%\label{tab:dataset}
%\begin{tabular}{|c|c|} \hline
%Name &Size(GB)\\ \hline \hline
%Text & $12,110$ \\ \hline
%Video & $246$ \\ \hline
%Software & $1,749$ \\ \hline
%\end{tabular}}
%
%\subtable[Real-Time Data Set]{
%\label{tab:dataset2}
%\begin{tabular}{|c|c|} \hline
%Name &Size(GB) \\ \hline \hline
%System log & $132$ \\ \hline
%User personal data & $1,651$ \\ \hline
%System backup image & $3,572$ \\ \hline
%\end{tabular}}
%\end{table}
%
%In order to evaluate the performances of \searsa under real-time application
%scenarios,
%
%We use the data set shown in Table~\ref{tab:dataset2} generated
We generate a data set reflecting real-time data access of 10 users during a span of 3
weeks in 2014 containing three parts.
(1)
%\textbf{User Personal Data}
User Personal Data
of 1.6 TB consisting
of various common types of files from 10 users;
(2)
%\textbf{System Log}
System Log
of 132 GB
consisting of major system log files (e.g. files under $/var/log$
directory) of Amazon EC2 Ubuntu server machines recorded every hour;
%The log files are recorded every hour.
and
(3)
%\textbf{System Backup Image}
System Backup Image
of 3.5 TB consisting of the complete backup
image files for Linux systems created once a day.

%\subsection{\textbf{Performance Metrics}}
%\label{subsec:metric}

%For each data set above, we evaluate \searsa
%in three dimensions: storage usage, time performance and
%reliability using the following metrics.

We evaluate \searsa in terms of storage usage with deduplication ratio and time performance with the average file retrieval time.
%
%using the following metrics.
%\begin{itemize}
%  \item \
\textbf{Deduplication Ratio} is defined as the ratio of the total
    size of original files over the total space consumption for SEARS
    including the indexing overhead for storing them. This metric
    captures the combined effect of deduplication (reduce space usage)
    and erasure coding (increase space usage).
%
%  \item
\textbf{Average File Retrieval Time} is defined as the average
    time duration from the moment the user issues a request for a file
    to the moment the file is ready at end device. This involves downloading and decoding of all necessary chunks
    and reconstruction of the file from all chunks.
% for systems using coding.

%\end{itemize}

%\subsection{\textbf{Evaluation Setup}}
%\label{subsec:setup}

%For the archival data set in Table~\ref{tab:dataset}, we directly feed the data
%into each storage system under evaluation.
%For the real-time data set in Table~\ref{tab:dataset2},

We employ 10 Amazon EC2 instances as driver machines to generate the log files, system backup images in addition to making users upload their own personal data.
We fix cluster size at $n=10$ thus use $10$ EC2 instances for each cluster. We use $E=20$ clusters.

We compare \searsa with
%two related storage systems
the existing storage system R-ADMAD~\cite{Liu-Rad-2009}
%using
%deduplication to reduce storage requirement.
%
%The Deep Store archival system~\cite{Bha-Pro-2006} chooses for each
%chunk a replication level that is a function of the amount of data
%that would be lost if that chunk were lost to increase data
%reliability. This results in an average replication level of $4.3$ in
%our date sets.
%
which packs variable-length data
chunks into fixed size objects of $8$ MB which are encoded with
erasure code and distributed among storage nodes called redundancy
groups.
To fairly evaluate
%these systems with
R-ADMAD with
\sears, we implement
%the two related work
it
%also
on EC2 cloud, and follow the same chunking process as \sears as specified in
%Section~\ref{subsec:chunking}
Section~\ref{sec:system}
for all files in our data
%sets
set
to generate chunks of 4 KB average size.
Furthermore, the same set of nodes are used for the \searsa cluster and the R-ADMAD
redundancy group.

%% two figures
%\begin{figure}[htbp]%
%\subfigure{\includegraphics[width=0.5\linewidth]{dedup_ratio-k_n_ratio.pdf}\label{fig:dedup_ratio-k_n_ratio}}%
%\subfigure{\includegraphics[width=0.5\linewidth]{retrieval_time-k_n_ratio.pdf}\label{fig:retrieval_time-k_n_ratio}}%
%\caption{(a) k/n effect on Dedup ratio; (b) k/n effect on retrieval time.}%
%\label{fig:firstSimFig}
%\end{figure}

%\begin{figure}[htbp]%
%\begin{figure}[tbp]%
%\subfigure{\includegraphics[width=0.5\linewidth]{dedup-realtime_dataset.pdf}\label{fig:dedup-realtime_dataset}}%
%\subfigure{\includegraphics[width=0.5\linewidth]{retrieval_time-no_IO.pdf}\label{fig:retrieval_time-no_IO}}
%\caption{(a)Dedup ratio; (b)File Retrieval time}
%\label{fig:secondSimFig}
%\end{figure}

%%% one figure for all results

\begin{figure*}[htbp]%
\subfigure{\includegraphics[width=0.23\linewidth]{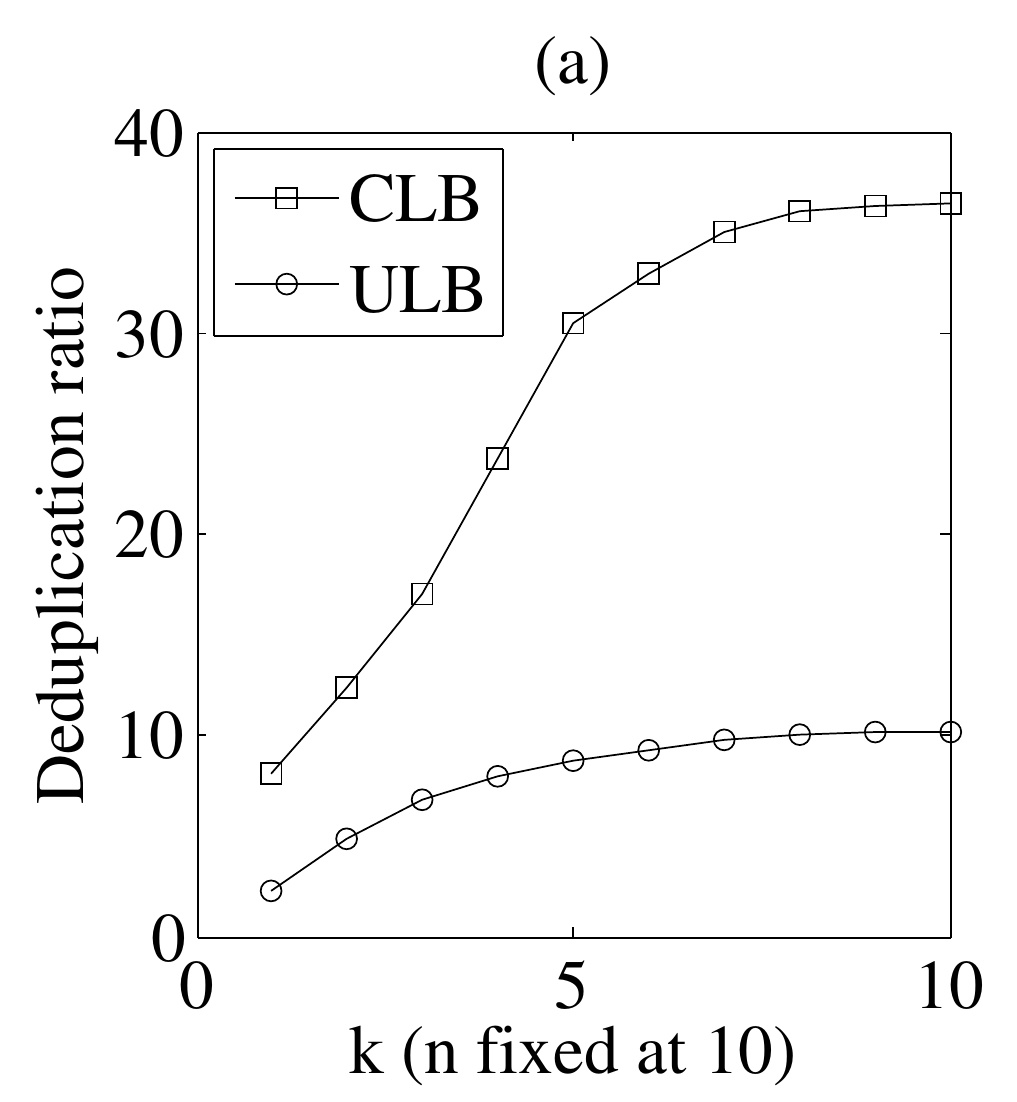}\label{fig:dedup_ratio-k_n_ratio}}%
\subfigure{\includegraphics[width=0.23\linewidth]{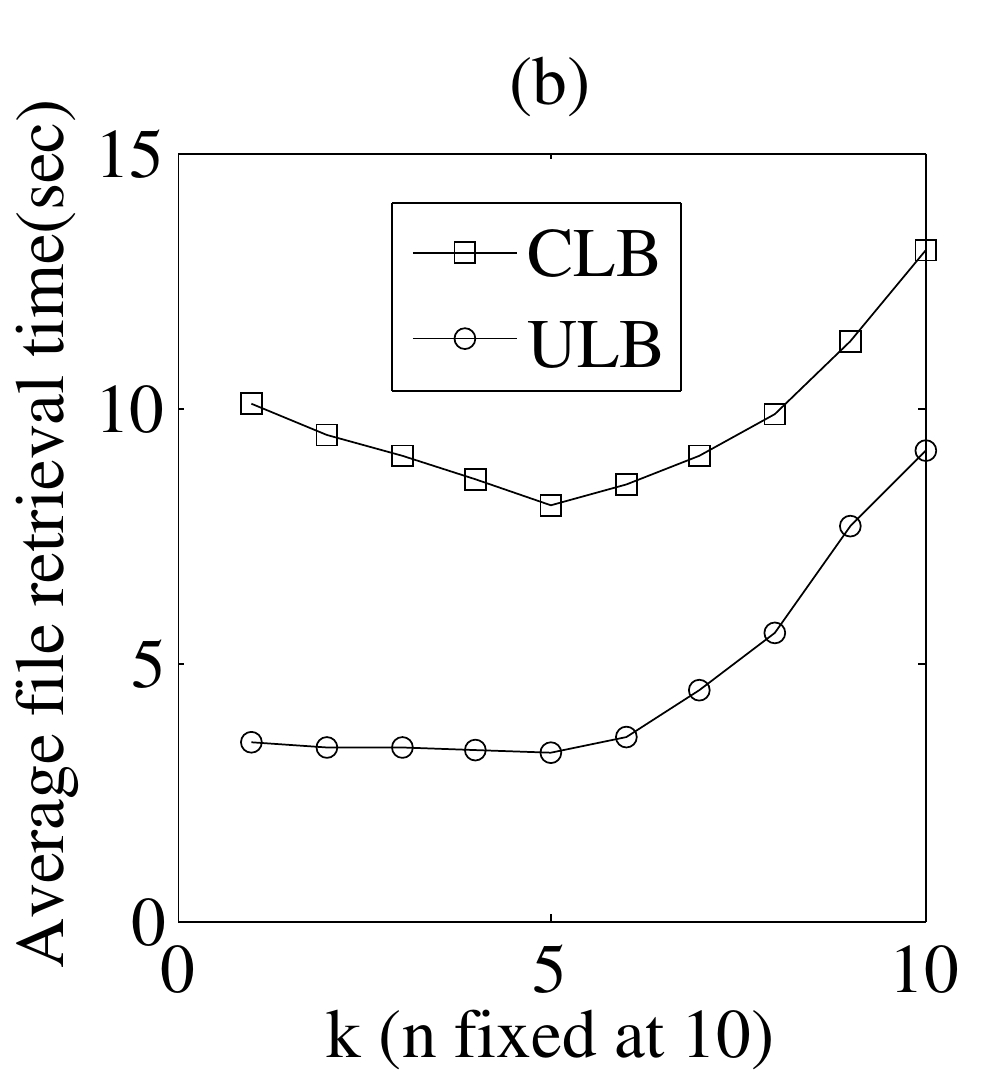}\label{fig:retrieval_time-k_n_ratio}}%
\subfigure{\includegraphics[width=0.213\linewidth]{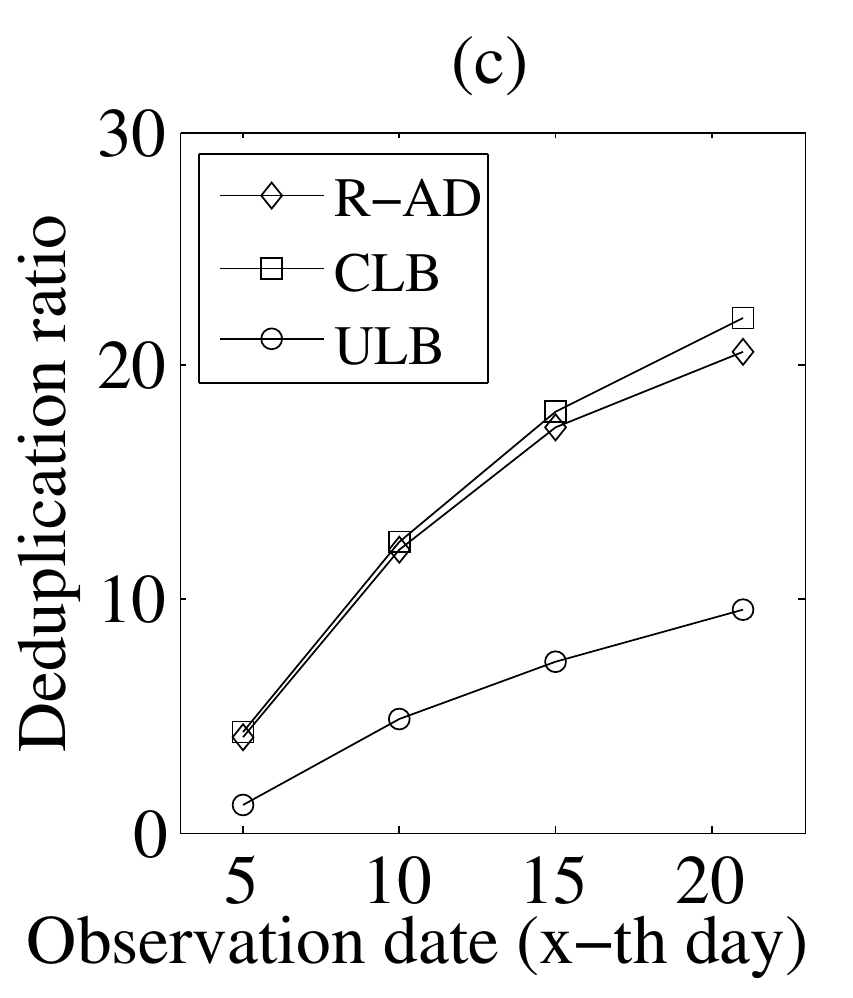}\label{fig:dedup-realtime_dataset}}%
\subfigure{\includegraphics[width=0.285\linewidth]{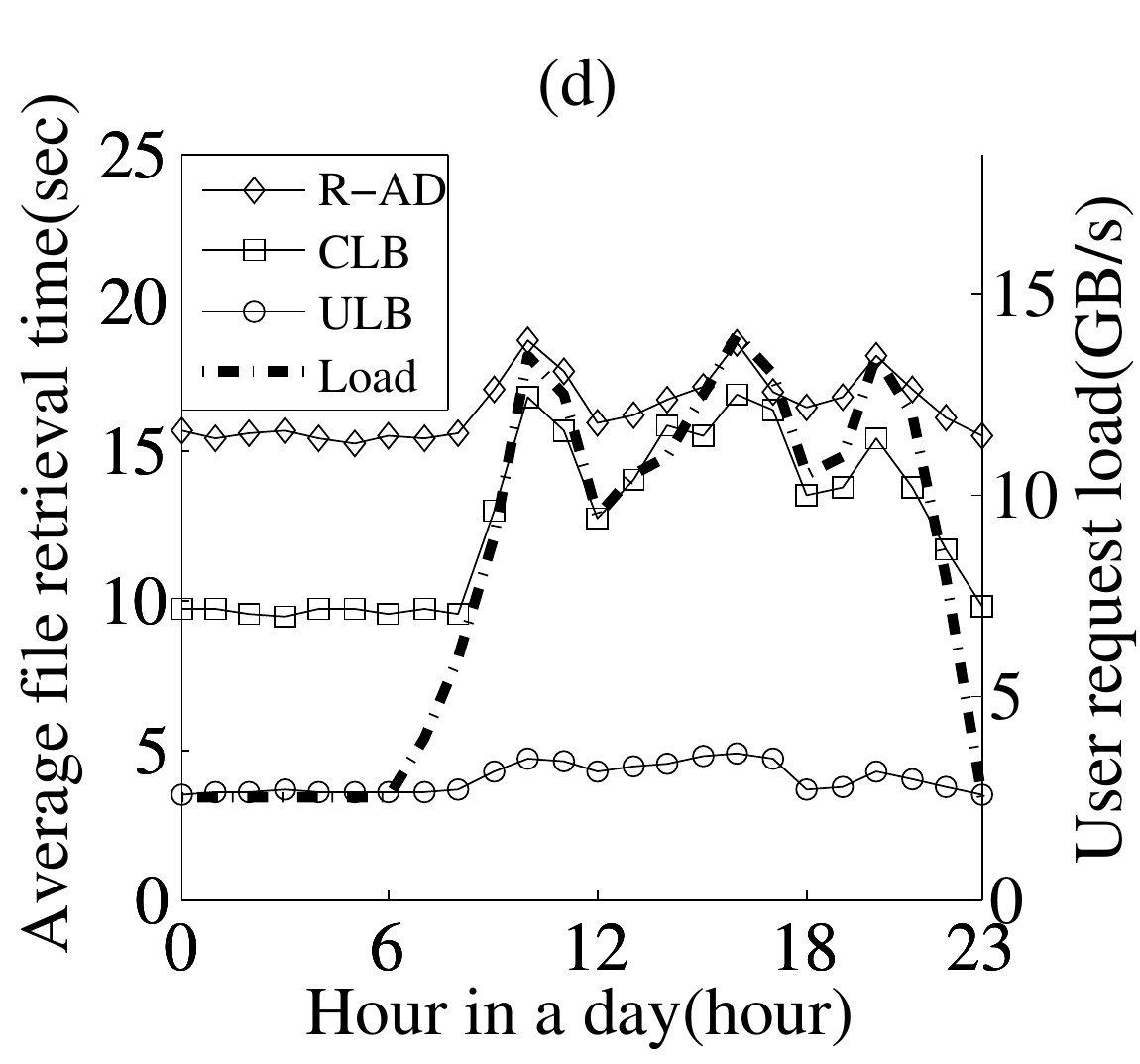}\label{fig:retrieval_time-no_IO}}
\caption{(a) k/n effect on Dedup ratio; (b) k/n effect on retrieval time; (c) Dedup ratio; (d) file retrieval time}
\end{figure*}

%\subsection{\textbf{Effect of k/n Ratio}}
%\label{subsec:ratio}
\textbf{Effect of k/n Ratio:}
The ratio $k/n$ has profound performance impact on any scheme using
erasure coding. To illustrate this, we fix $n$ at $10$ and vary $k$
for the
%combined
data
%sets.
set.
As each chunk requires $n/k$ times as much space as before the coding
process, deduplication ratio increases with $k$ as shown in
%Figure~\ref{fig:firstSimFig}-(a).
Figure~\ref{fig:dedup_ratio-k_n_ratio}.
Increases of $k$ also lead to larger numbers of code pieces with
smaller sizes for each chunk. This implies more parallel retrieval
processes, each with smaller bandwidth requirement. With smaller $k$
($k<5$), both factors contribute to reduced chunk and file retrieval
time. However, after $k$ increase beyond a threshold, $k=5$ for the
data sets, the larger number of concurrent retrieval processes and the
decoding process with more code pieces become the bottleneck and
increase retrieval time as shown in
%Figure~\ref{fig:firstSimFig}-(b).
Figure~\ref{fig:retrieval_time-k_n_ratio}.
CLB exploits redundancy across all chunks in all files and achieves
a higher deduplication ratio. However, the process of searching for
chunks across all clusters leads to the higher file retrieval
time. On the other hand, ULB can only exploit intra-user redundancy
which leads to a lower redundancy ratio. However all chunks in a
file are easily retrieved from one cluster, which leads to the faster
file retrieval time.
%FLB balances the tradeoff between the two.
%
%For the rest of the section,
%Next,
We use $k=5$ and $n=10$ from now on.
%when not specified.

%\subsection{\textbf{Storage Efficiency}}
%
%\subsubsection{\textbf{Deduplication Ratio}}
\textbf{Deduplication Ratio:}
%
%We measure deduplication ratio on both data sets and observe similar
%trend, that is,
%
%CLB, R-ADMAD, FLB, Deep Store, ULB.
%We show results from the archival set in
%Figure~\ref{fig:secondSimFig}-(a).
%
To see how the ratio changes as data volume evolves over time,
%for the real-time data set,
we plot the cumulative deduplication ratio on the
$5^{th}$, $10^{th}$, $15^{th}$, and $21^{st}$ day in
%Figure~\ref{fig:secondSimFig}.
Figure~\ref{fig:dedup-realtime_dataset}.
The ratio improves for all schemes
over time as data volume increases, for more redundancy can be exploited.
It also shows deduplication ratio
decreases in the order of CLB, R-ADMAD, and ULB.
R-ADMAD is essentially same as CLB in data deduplication as it can exploit system wide redundancy just as CLB.
But R-ADMAD uses slightly more space than CLB because of its indexing structure is more complex than CLB.

%\subsection{\textbf{Time Performance}}

%\subsubsection{File Retrieval Time}
%\label{subsec:time}
\textbf{Time Performance:}
To examine interactive user experience,
% with all proposed systems,
we
replay the request pattern captured
in the user personal data trace of
our
%the real-time
data set.
We use $10$ desktop machines residing in the eastern region of the US.
Each desktop
replays the file
access trace for each of the $10$ users.  We report the file
%uploading and
retrieval
time for files accessed during each hour of the day averaged over $21$
days over $10$ users.
%
%\subsubsection{\textbf{File Retrieval Time}}
%\label{subsec:time}
%To measure file retrieval time in its purist form, we assume all file chunk-meta-data are synchronized on the user's end device
%avoiding the interaction with the switching node.
To retrieve a file,
the user's end device directly requests data chunks
%from the node
%storing one replica of the chunk in Deep Store, and
from $10$ nodes
storing the code pieces of each chunk in the
%other four
three schemes.
%using erasure code.
%
Figure~\ref{fig:retrieval_time-no_IO} presents file retrieval time in relation to user request load averaged over each hour of the day over 21 days. Users' data request volume per hour in these figures reflect work activity
during a day, that is, light activity at night (0:00 midnight to 8:00 am) and heavy and fluctuating activity for the rest of the day.
ULB offers the fastest and relatively flat retrieval time because
requests from the same user are handled by one cluster and there are no
multiple requests for the same data chunk at the same time.  CLB
offers slower file retrieval than ULB,
%out of the three schemes in \sears,
and large fluctuation during the working hours closely matching data
request volume. This is because a unique chunk is stored only once in
the entire system, and multiple users can request the same unique
chunk at the same time, which leads to congestion at the cluster
hosting the chunk in demand.
%
%The retrieval time for FLB follows the
%same pattern but is smaller than CLB. This is the result of less
%number of concurrent requests being issued for the same unique chunk
%as all chunks of the same file must reside on the same cluster,
%therefore an identical chunk can appear in multiple clusters one for a
%different file.
%
%Both R-ADMAD and Deep Store follow the data volume fluctuation during
%the day but with larger retrieval time than the three \searsa schemes.
%
R-ADMAD follows the data volume fluctuation during
the day but with larger retrieval time than \sears.

To compare with a commercial system, we note that downloading $3$ MB
files from the same set of 10 desktops residing in the eastern part of
the US takes an average of $7$ s from Amazon EC2 service in us-east-1
region~\cite{EC2Measure}. With ULB in \sears, the download time is
$2.5$ s throughout the day.

\section{Conclusion and Future Work}
\label{sec:conc}
%In this work, we proposed

We describe the design and implementation of a space efficient, data
reliable and fast retrieving cloud-based storage system \searsa which
integrates data deduplication and erasure coding. \searsa
provides a flexible combination of various
%storage server
binding schemes to associate server nodes with
data to be stored at different level based on
%different
application
needs.
%The use of erasure coding allows for flexible distribution of code pieces to different
%nodes, leading to increased chance of reducing storage space through
%data deduplication and load-balanced node usage,
%as well as enabling
%concurrent and fast file retrieval.
%
%Performance
Evaluation
%studies
over Amazon
EC2 shows that \searsa
%significantly
outperforms related systems with lower storage usage while ensuring fast and reliable data access.

As future work, we plan on examining the location of cluster nodes
inside data centers to future improve data reliability and reduce
retrieval time.  We are evaluating the system with more data sets with
additional metrics such as storage balance, file upload time and file
retrieval success rate.  Various system design parameters in \searsa
and performance under flexible configuration of \searsa with multiple
binding schemes, chunk size and erasure codes also need further
investigation.

\bibliographystyle{abbrv}
\bibliography{references,coding}
%

%\balancecolumns
\end{document}